\documentclass[fleqn,12pt,twoside]{article}
\usepackage{espcrc1}
\usepackage{graphicx}
\usepackage{wrapfig}

\usepackage[figuresright]{rotating}

\newcommand{\AmS}{{\protect\the\textfont2
  A\kern-.1667em\lower.5ex\hbox{M}\kern-.125emS}}

\title{Leptonic Observables in Ultra-Relativistic Heavy Ion Collisions}

\author{J.L. Nagle\address[Columbia]{Columbia University, New York, NY 10027, USA} for the PHENIX Collaboration{\thanks{for the full PHENIX Collaboration author list and acknowledgements, see Appendix "Collaborations" of this volume.}}}

\begin{document}

% typeset front matter
\maketitle

\begin{abstract}
We report on leptonic observables by the PHENIX
experiment from data taken during Run II at the Relativistic Heavy Ion Collider (RHIC).  
We show first results on $\phi \longrightarrow K^{+}K^{-},e^{+}e^{-}$, low and
intermediate mass dielectron continuum, single electrons from charm, and $J/\psi$ 
yields in proton-proton and Au-Au collisions at $\sqrt{s_{NN}}=200$ GeV.
\end{abstract}

\section{PHENIX Experiment at RHIC}
The PHENIX experiment was designed with the specific capability of sampling
high luminosity proton-proton, proton-ion and ion-ion collisions.  This feature,
combined with a suite of detectors for measuring both electrons in two central arm 
spectrometers and muons in two forward arm spectrometers, allows for a large
range of interesting measurements that probe the earliest stages of the hot
quark-gluon matter created at RHIC.  Further details of the detector design and
performance are given in~\cite{phenix-nim}.

The PHENIX experiment consists of two central spectrometer arms that were
completely instrumented in Run II at RHIC.  The central arms each cover
pseudorapidity ($|\eta|<0.35$), transverse momentum ($p_{T}>0.2$ GeV/c), 
and 90 degrees in azimuthal angle $\phi$.  They are comprised from the inner radius outward
of a Multiplicity and Vertex Detector (MVD), Drift Chambers (DC), Pixel Pad Chambers (PC),
Ring Imaging Cherenkov Counter (RICH), 
Time Expansion Chamber (TEC), Time-of-Flight Scintillator Wall (TOF), 
and two types of Electromagnetic Calorimeters (EMC).
This combination of detectors is necessary for the clean identification of
electrons over a broad range in transverse momentum.  In addition, we have
excellent photon capabilities with the EMC and high precision hadron identification
with our smaller coverage TOF system.

PHENIX has two forward muon spectrometers consisting of hadronic absorbers,
a cathode strip chamber muon tracking system, and interleaved Iarocci tubes with
steel plates for muon identification and triggering.  Each spectrometer covers
approximately $1.2 < |\eta| < 2.2$ and $p_{tot}>2$ GeV/c.  
The south muon spectrometer was commissioned and
operational in Run II, while the north arm is currently being completed and will
be ready for data taking in Run III.

Au-Au and proton-proton collisions are characterized with minimum bias triggers based
on a set of zero degree calorimeters (ZDC), beam-beam counters (BBC), and a scintillator 
multiplicity counter (NTC) for larger coverage in proton-proton collisions.  

\section{Run II at RHIC}

\begin{figure}
\begin{center}
% note that 30pc centered well
\includegraphics[width=15pc]{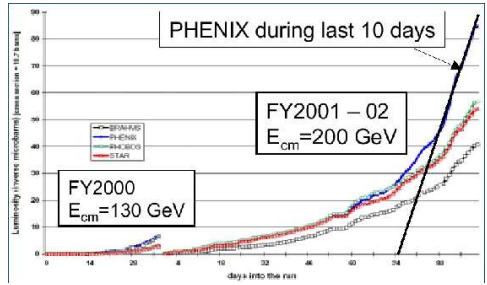}
\caption{Integrated Au-Au luminosity delivered by the RHIC accelerator to each of the
four experimental intersection regions as a function of days into the run.}
\label{fig_luminosity}
\end{center}
\end{figure}

During the second running period at RHIC the
accelerator achieved full design energy $\sqrt{s_{NN}}=200$ GeV for Au-Au collisions.  
RHIC delivered to the PHENIX intersection region
approximately 42 $\mu{\rm b}^{-1}$ within a vertex $z$ range ($|z|<45$~cm). PHENIX
was successful in sampling this luminosity with a combination of ``minimum
bias'' triggers and a full complement of software based Level-2 triggers.  
As shown in Figure~\ref{fig_luminosity}, over 50\% of the integrated luminosity 
was delivered during the last two weeks of the Au-Au running period.  Thus,
we are seeing just the beginning of high luminosity running at RHIC.  In the
analysis shown in these proceedings, we have scanned of order 26 million ``minimum
bias'' events with $|z|< 30$~cm.  

The RHIC accelerator also commissioned polarized proton-proton collisions and had 
five weeks of experiment data taking.  
The collisions were at $\sqrt{s}=200$ GeV for optimal comparison with the Au-Au data set.  The
PHENIX experiment commissioned a first set of fully pipelined hardware Level-1 triggers including 
triggers on high energy photons, single electrons, single muons, and dielectrons
and dimuons from $J/\psi$.  We observed with ``minimum bias'' and Level-1 
triggers approximately 150~nb$^{-1}$. 

\section{$\phi$ Meson}

The $\phi(1020)$ vector meson is particularly interesting because
the restoration of approximate chiral symmetry at high
temperature may modify the $\phi$ mass and width~\cite{phi}.  These modifications may result
in a change in the branching fraction of $\phi \longrightarrow K^{+}K^{-}$
and $\phi \longrightarrow e^{+}e^{-}$ when the $\phi$ decays in medium.  Note
that the $\phi$ lifetime $\tau \approx 44$ fm/c is longer than the expected lifetime
of the coupled collision system, and thus only a fraction may decay in the hot
fireball.
It has also been hypothesized that final state interactions of kaons from $\phi$
decay may lower the apparent measured branching fraction in the kaon channel~\cite{johnson}.
PHENIX has the unique ability to reconstruct the $\phi$ meson in both the 
$K^{+}K^{-}$ and $e^{+}e^{-}$ channels.  

\begin{figure}[htb]
\begin{minipage}[t]{80mm}
\includegraphics[width=20pc]{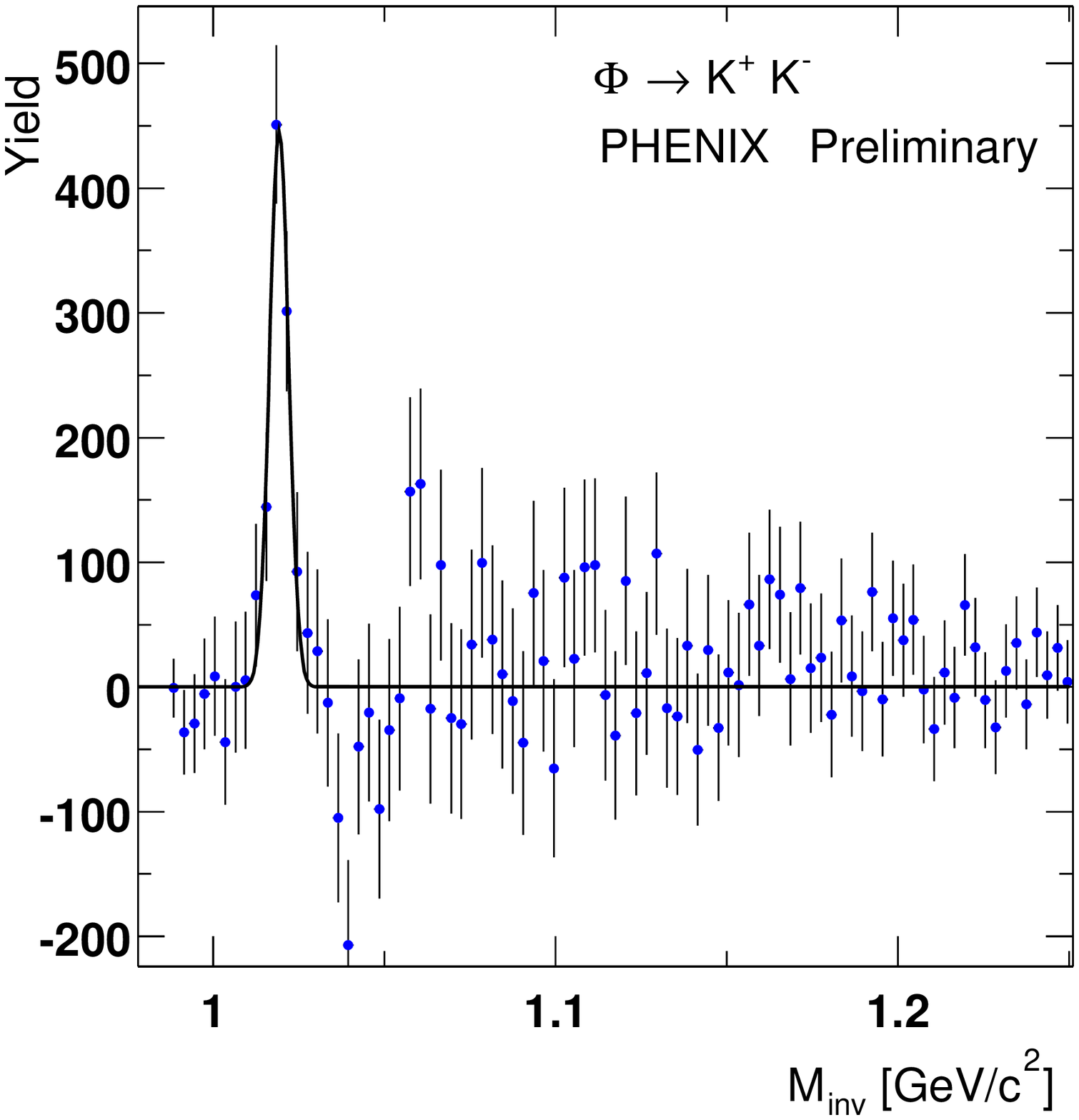}
\caption{$K^{+}K^{-}$ invariant mass distribution after mixed event subtraction
for minimum bias (0-90\% central) Au-Au collisions at $\sqrt{s_{NN}}=200$~GeV.}
\label{fig:phi_kaons}
\end{minipage}
\hspace{\fill}
\begin{minipage}[t]{75mm}
\includegraphics[width=20pc]{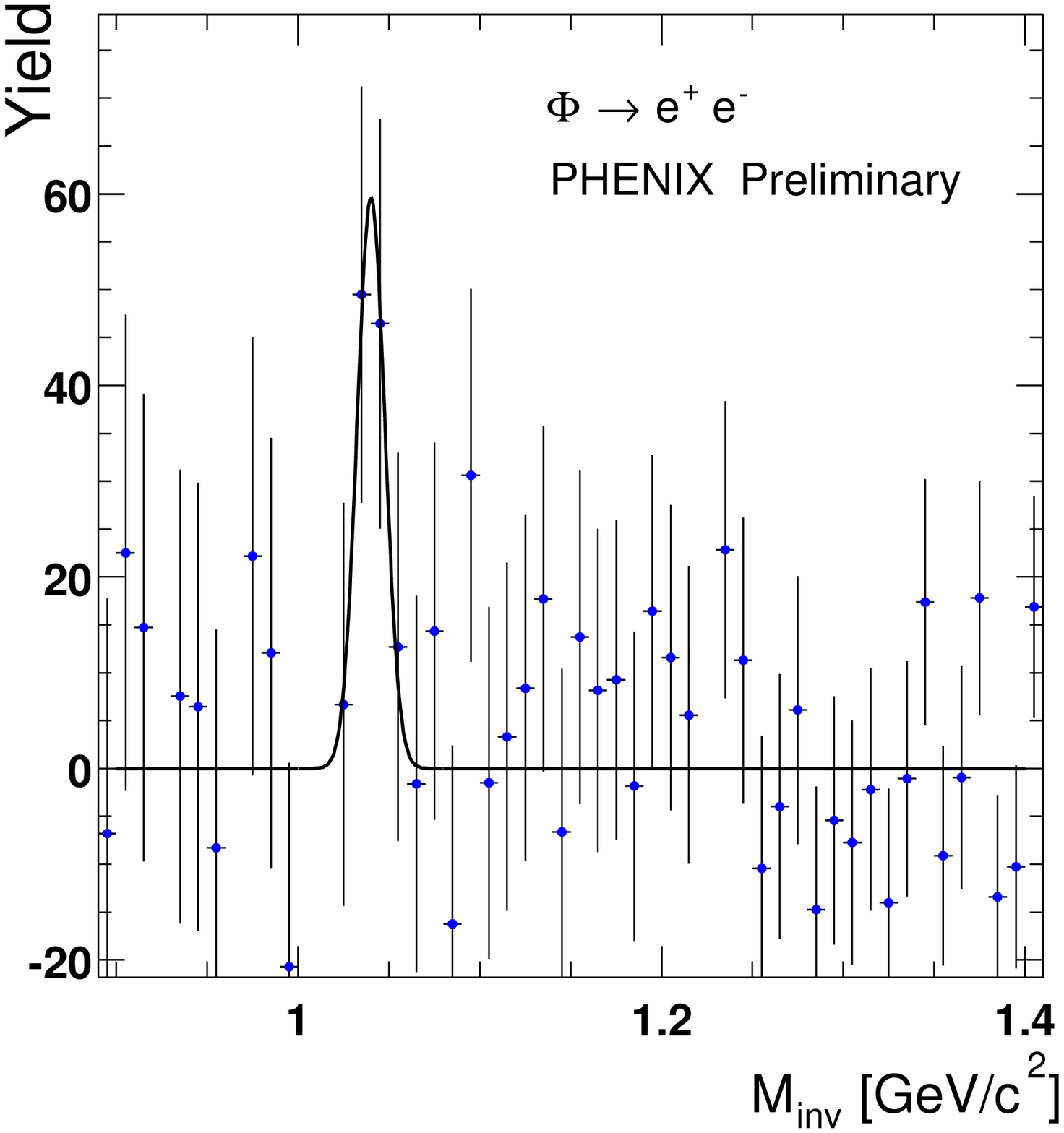}
\caption{$e^{+}e^{-}$ invariant mass distribution after mixed event subtraction
for minimum bias (0-90\% central) Au-Au collisions at $\sqrt{s_{NN}}=200$~GeV.}
\label{fig:phi_electrons}
\end{minipage}
\end{figure}

PHENIX has excellent kaon particle identification using our high precision
time-of-flight scintillator wall.  Shown in 
Figure~\ref{fig:phi_kaons} is the $K^{+}K^{-}$ invariant mass distribution
after mixed event background subtraction.  The distribution is for 
minimum bias (0-90\% central) Au-Au collisions at $\sqrt{s_{NN}}=200$ GeV.  
The signal to background ratio is 1/12 and the mass peak and width values agree
within errors with the Particle Data Group values.  
More details
are shown in~\cite{debsankar}.  
After correcting for acceptance,
efficiencies, and the branching fraction in vacuum $\phi \longrightarrow K^{+}K^{-}$ 
(0.49), we find:
\begin{equation}
\phi~({\rm via~K^{+}K^{-}}):~~{{dN}\over{dy}}|_{y=0}=2.01 \pm 0.22({\rm stat}) ^{+1.01}_{-0.52}({\rm sys})~~~{\rm [PRELIMINARY]}
\end{equation}
%The systematic error includes contributions for the current uncertainty in the 
%transverse momentum slope and the extrapolation beyond our acceptance.
This value is in agreement with the STAR experiment results reported at
this conference~\cite{geno} within errors.
We have higher statistics using time information from our larger coverage 
electromagnetic calorimeter for particle identification.  We expect to have 
transverse mass distributions for a full range of Au-Au centrality bins shortly.

PHENIX has excellent electron identification capabilities that are necessary
to separate electrons from the much more abundant charged pions.  The RICH 
yields a threshold selection for electrons, the TEC allows for pion-electron 
separation at low momentum due to different energy loss $dE/dx$, and the EMC is used
to verify a tracking momentum-electromagnetic energy match.  We show in 
Figure~\ref{fig:phi_electrons} the $e^{+}e^{-}$ invariant mass distribution
after mixed event background subtraction.  There is an excess of counts at
the $\phi$ mass with a signal strength of $101 \pm 47 ({\rm stat}) ^{+56}_{-20}({\rm sys})$
and a signal to background ratio of 1/20.    Within relatively large errors
the mass peak and width values agree with the values from the
Particle Data Group.  Using the signal value quoted above, we have corrected
for the acceptance and efficiencies, and assuming the branching
fraction in vacuum $\phi \longrightarrow e^{+}e^{-}$ = $2.9 \times 10^{-4}$, we find:
\begin{equation}
\phi~({\rm via~e^{+}e^{-}}):~~{{dN}\over{dy}}|_{y=0}=5.4 \pm 2.5({\rm stat}) ^{+3.4}_{-2.8}({\rm sys})~~~{\rm [PRELIMINARY]}
\end{equation}
Although the electron and kaon results differ, they are consistent within 
$1~\sigma$ statistical errors.  
We need further statistics in the electron channel from future running.

\section{Dielectron Continuum}

In the dielectron invariant mass region below the
$\phi$, referred to as the Low Mass Region (LMR), there may be excess dielectrons
from in-medium mass modification of the $\rho$ meson due to restoration
of approximate chiral symmetry or collision broadening~\cite{chiral-rapp}.  
In the region above the $\phi$ and below the $J/\psi$, referred to as the Intermediate Mass Region 
(IMR), there may be a significant contribution from semi-leptonic decays of
open charm D mesons at RHIC energies.

\begin{figure}[htb]
\begin{minipage}[t]{80mm}
\includegraphics[width=20pc]{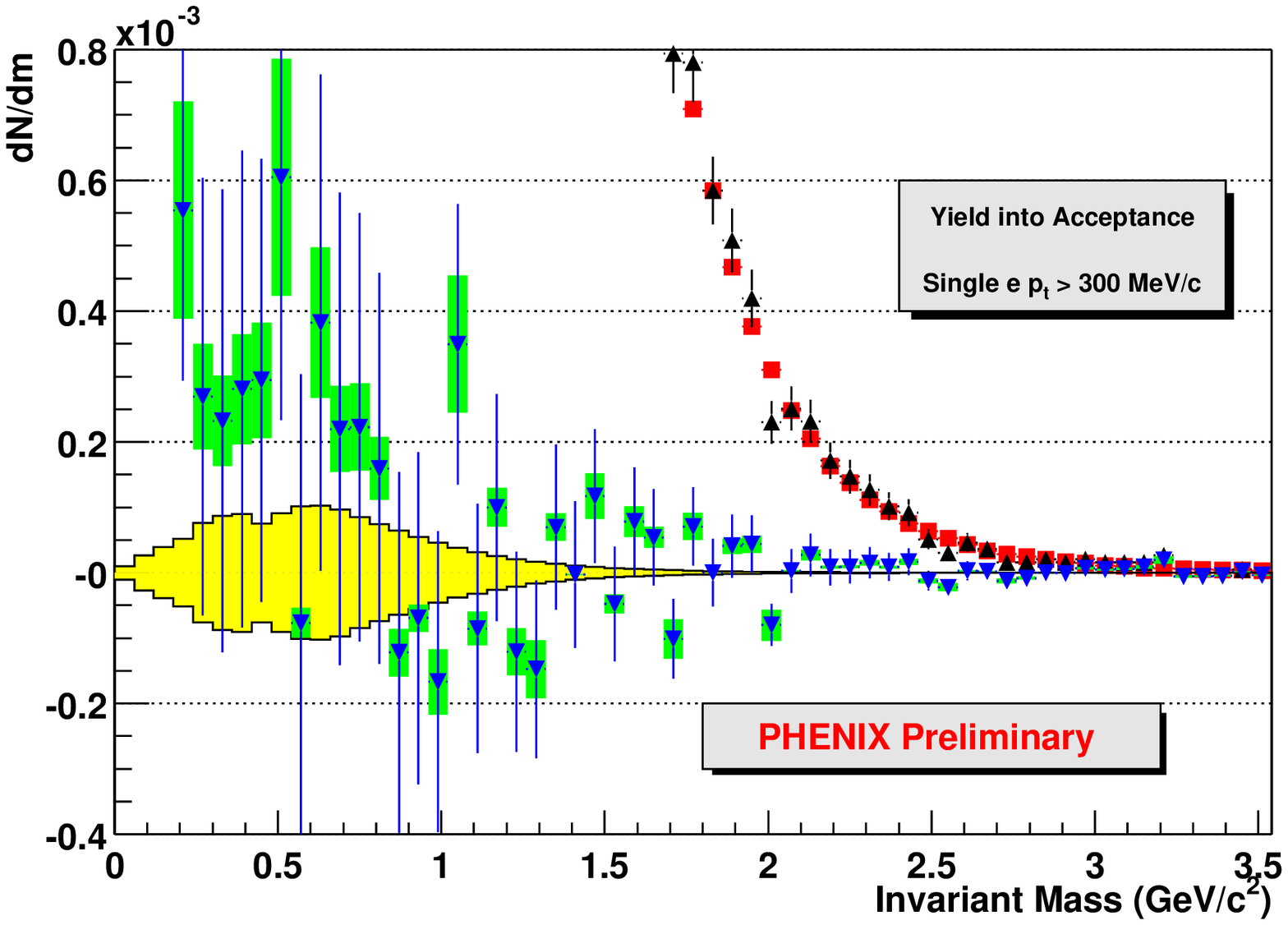}
\caption{dN/dm ($c^2/GeV$) into the PHENIX acceptance in ``minimum bias'' Au-Au events for 
dielectrons as a function of invariant mass after mixed event
background subtraction.}
\label{fig-dielectron-sys}
\end{minipage}
\hspace{\fill}
\begin{minipage}[t]{75mm}
\includegraphics[width=20pc]{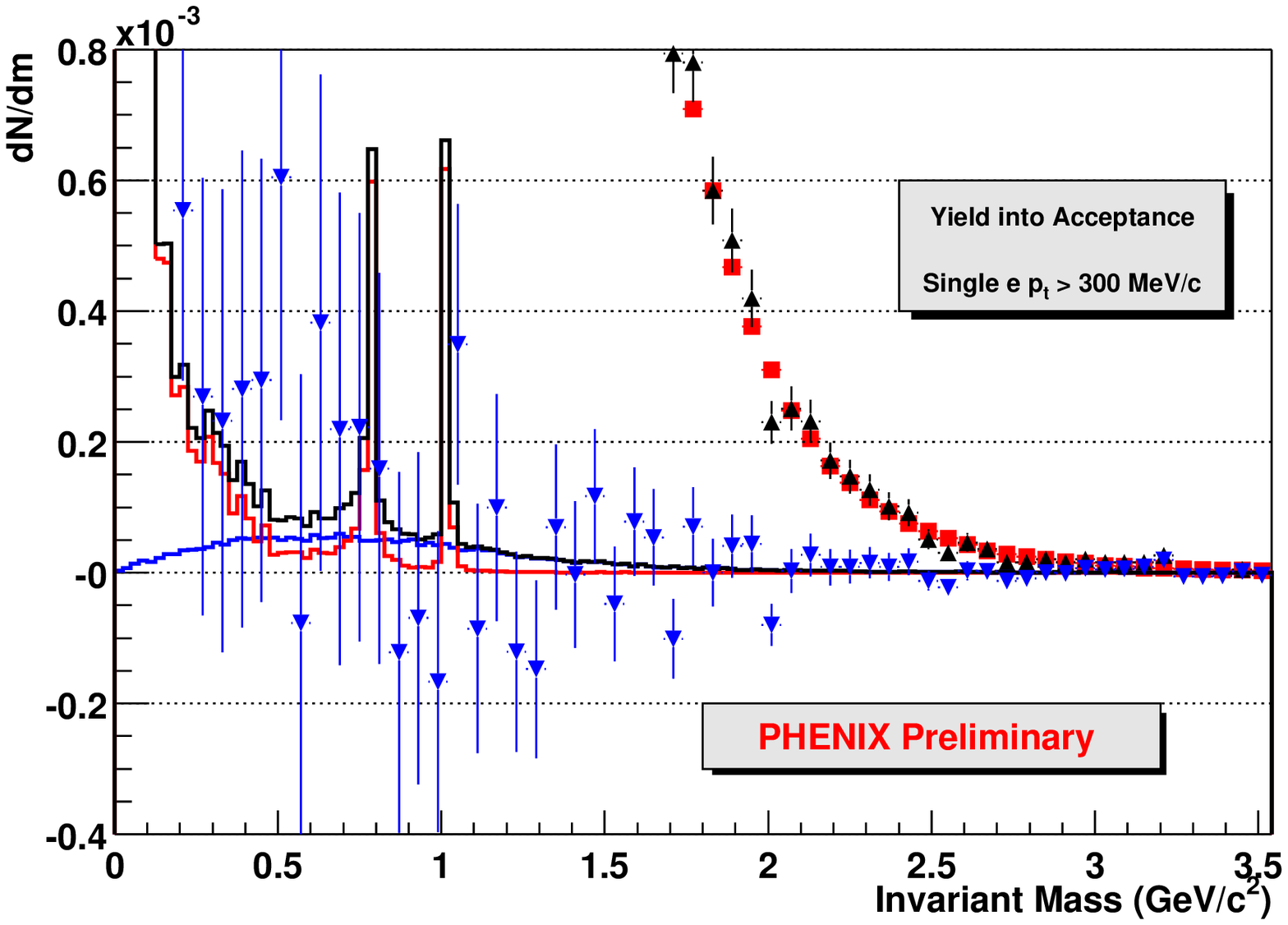}
\caption{Same data as in left figure without systematic error bands.  
Data is overlayed with ``baseline'' expectations calculation assuming no medium modifications.}
\label{fig-dielectron-model}
\end{minipage}
\end{figure}

We show in Figure~\ref{fig-dielectron-sys} dN/dm into the PHENIX acceptance as a
function of dielectron invariant mass after
mixed event background subtraction with statistical and systematic errors.  
Note that we have corrected for our efficiencies relative to our PHENIX specific
acceptance, and will publish our simple acceptance definition shortly.
The solid band represents the systematic error in the normalization 
of the mixed event background, while the
individual bands are the point to point systematics.  We show the same distribution in 
Figure~\ref{fig-dielectron-model} with a simple ``cocktail'' model of expected  
baseline contributions (black), including light meson decays (red) and  
a PYTHIA charm calculation scaled by the number of binary collisions (blue).

In the LMR ($M_{inv}=0.3-1.0$~GeV/c$^{2}$), the dominant expected contributions are 
from meson Dalitz decays and low mass vector meson decays (including the $\omega$).  
Assuming no medium modifications we
expect a yield of $N = 9.2 \times 10^{-5}$~(counts/event).  The PHENIX preliminary result is:
\begin{equation}
{{N}}~({\rm LMR,~into~PHENIX})=13.4 \pm 7.2({\rm stat})^{+12.2}_{-8.4}({\rm sys}) \times 10^{-5}~[{\rm counts/event}]
\end{equation}
The large statistical and systematic errors prevent an observation of the $\omega$ meson or conclusions
about $\rho$ mass shifts at this time.   

In the IMR ($M_{inv}=1.1-2.5$~GeV/c$^{2}$), the dominant expected contribution is from
the correlated semi-leptonic decays of D and associated $\overline{D}$ mesons.  
Using a PYTHIA charm calculation, assuming binary collision scaling for ion-ion reactions, 
we would expect $N = 1.5 \times 10^{-5}~({\rm counts/event})$.  
Our preliminary result in the IMR is:
\begin{equation}
{{N}}~({\rm IMR,~into~PHENIX})=0.38 \pm 2.60({\rm stat})^{+1.40}_{-0.81}({\rm sys}) \times 10^{-5}~[{\rm counts/event}]
\end{equation}
This result is consistent with the charm expectation within errors, but clearly requires further
statistics to demonstrate a signal.  
PHENIX has many handles on the production of heavy quarks, and the first results
are via the measurement of single electrons.  

\section{Charm via Single Electrons}

The measurement of open charm and open beauty in heavy ion collisions 
is sensitive to the initial gluon densities in the incoming nuclear 
wavefunctions and possible thermal contributions from later gluon-gluon scattering.  In
addition, total charm production is a crucial comparison by which to gauge
the possible suppression of $J/\psi$ in deconfined matter.  
Total charm production (integrated over all
transverse momentum) may be expected to be 
suppressed relative to binary scaling due to shadowing of the gluon parton distribution in nuclei or
enhanced due to thermal production in later stages.  Also, it was first predicted that
high transverse momentum charm quarks will lose energy in a dense gluonic medium, and then
later postulated that this energy loss is tempered by the ``dead-cone'' effect that
suppresses co-linear gluon emission for $v<<c$ partons~\cite{dima}.  Also, if the charm mesons
appear in medium they may undergo hadronic interactions.

Electrons resulting from semi-leptonic decays of charm D mesons ($D \longrightarrow
e + K + \nu$) and beauty B mesons ($B \longrightarrow e + D + \nu$) 
have significant contributions to inclusive
measured electrons above 1.0 and 2.5 GeV, respectively.  The dominant sources of
``background'' electrons are from neutral pion Dalitz decay ($\pi^{0} \longrightarrow
\gamma e^{+}e^{-}$) and photon conversions in material in the PHENIX aperture.  
We can account for these sources using our measured $\pi^{0}$ distributions.  In addition,
although their contributions are relatively small, we must model additional electrons 
from $\eta, \eta', \omega, \phi, \rho$.  

\begin{figure}[htb]
\begin{center}
\includegraphics[height=16pc]{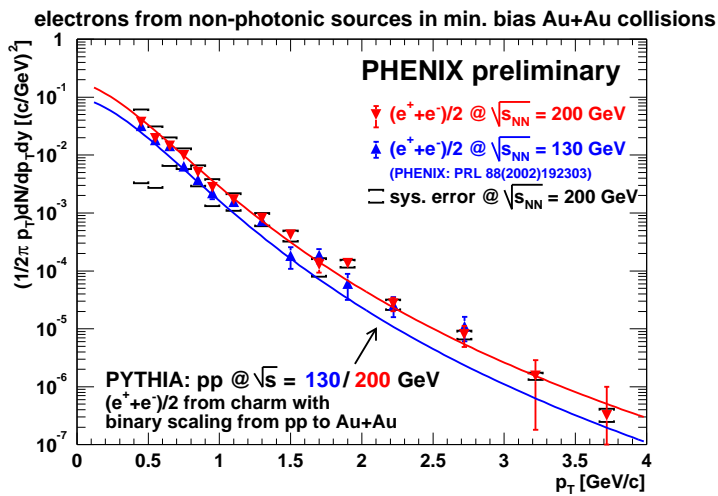}
\caption{Invariant transverse momentum distribution of electrons from 
non-photonic sources in ``minimum bias'' Au-Au collisions compared 
with expectations from PYTHIA for semi-leptonic charm D meson decay.}
\label{fig-elec}
\end{center}
\end{figure}

We have published~\cite{phx-e} our results from Run I Au-Au collisions at $\sqrt{s_{NN}}=130$ GeV and find a 
significant excess of electrons over ``background'' above $p_{T} \approx 0.7$ GeV/c that increases 
with $p_{T}$.  
The lower data points in Figure~\ref{fig-elec} are the minimum bias transverse 
momentum distribution of non-photonic electrons.
The results are consistent within relatively large errors with a 
PYTHIA 6.152~\cite{pythia} charm calculation using CTEQ5L parton distribution
functions and assuming that charm scales with the number of binary collisions.  

The production mechanism within PYTHIA is dominantly the
fusion of gluons to produce a $c\overline{c}$ pair.  The charm quark then fragments
into a D meson which can decay semi-leptonically yielding an electron.  However, an
alternative picture is that the charm quarks re-scatter in the medium and take part
in hydrodynamic expansion.  We have used PHENIX derived hydrodynamic
model parameters from our measured pion, kaon and proton data, and 
calculated the shape of the D meson distribution.  After simulating the D meson decays,
we find equally good agreement with our Run I single electron transverse momentum shape.

In Run II, we have a more powerful method of extracting the single electron charm
contribution.  Most of the ``background'' sources are photonic in origin.  Thus,
in Run II we had a special set of runs with a brass photon converter wrapped around the
beam pipe near the interaction point.  This material increases by a fixed factor the
number of electrons whose source is photonic.  Since there is a fixed relation between
the $\gamma$ from $\pi^{0}$ decay (which is the dominant source of conversion photons),
and the Dalitz contribution $\pi^{0} \longrightarrow \gamma e^{+}e^{-}$, we can 
subtract out both contributions.  This method has many advantages over the previously
described ``cocktail'' subtraction method, and has completely independent systematic
errors.  Also, this method subtracts out any direct photon contribution to 
conversion electrons, while the ``cocktail'' method does not.

After subtraction of the photonic contributions, we
show as the upper points in Figure~\ref{fig-elec} the minimum bias transverse momentum distribution of
non-photonic electrons.  Again, the data appear reasonably described by a PYTHIA calculation
of the expected charm contribution assuming binary collision scaling.  The electron
data in four centrality bins appear to qualitatively follow binary collision scaling
as described in~\cite{ralf}.  Significant work remains to reduce the systematic errors
in order to exactly quantify this scaling relation.  PHENIX is currently analyzing the
proton-proton data set so that we can calculate the charm scaling without relying
on the PYTHIA baseline.

\begin{figure}
\begin{minipage}[b]{0.33\linewidth}
\centering
\includegraphics[width=1.0\textwidth]{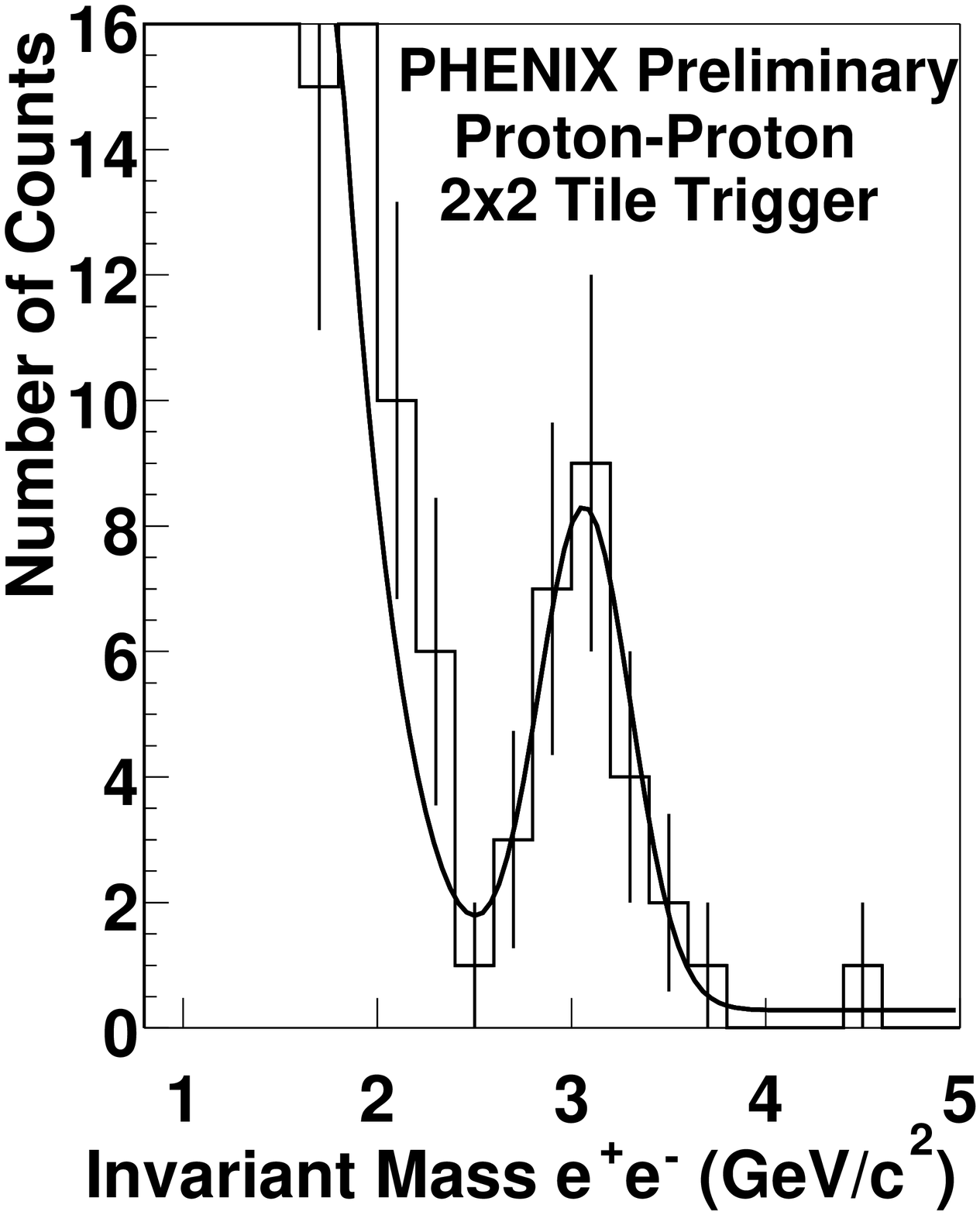}
\end{minipage}%
\begin{minipage}[b]{0.33\linewidth}
\centering
\includegraphics[width=1.0\textwidth]{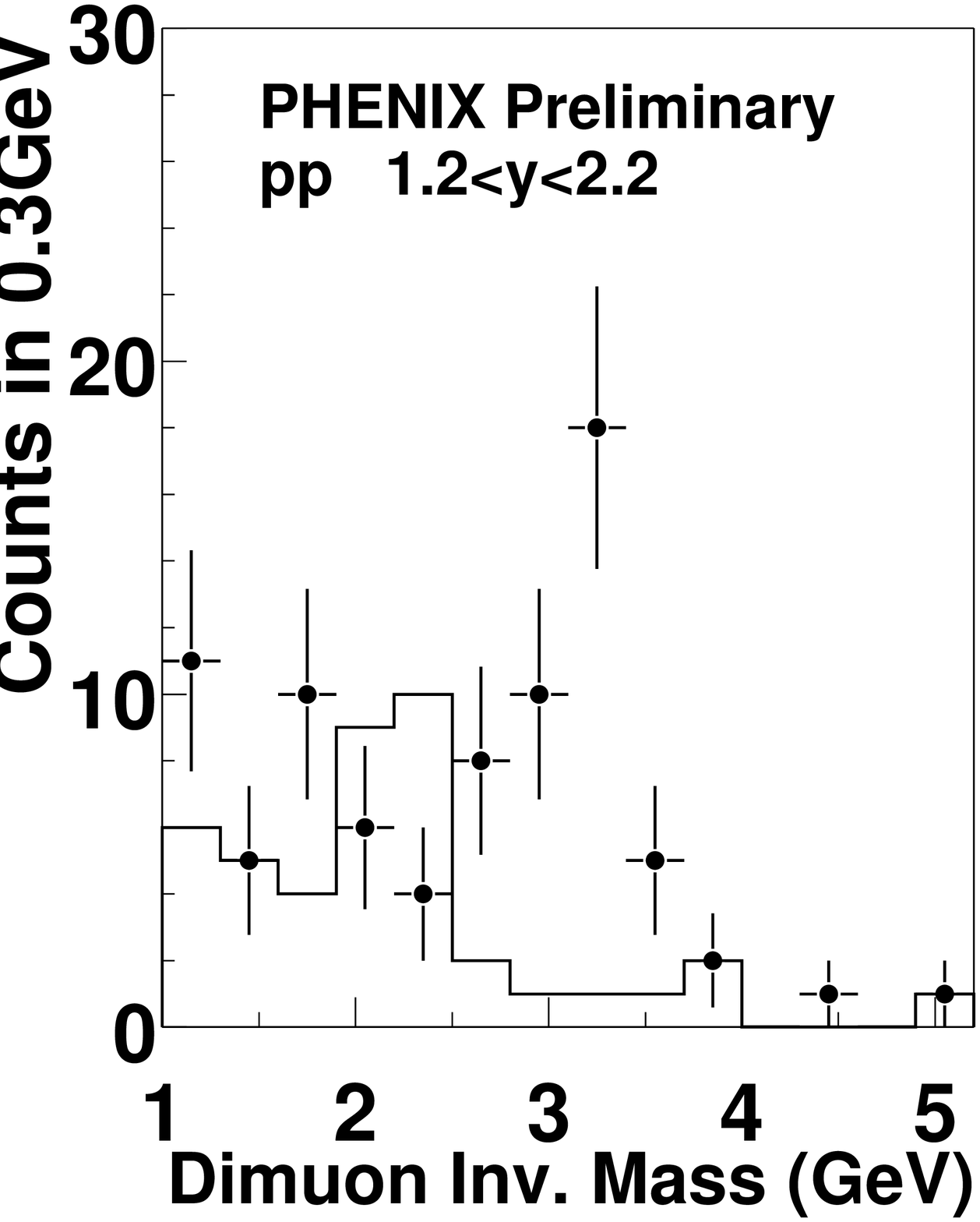}
\end{minipage}%
\begin{minipage}[b]{0.33\linewidth}
\centering
\includegraphics[width=1.0\textwidth]{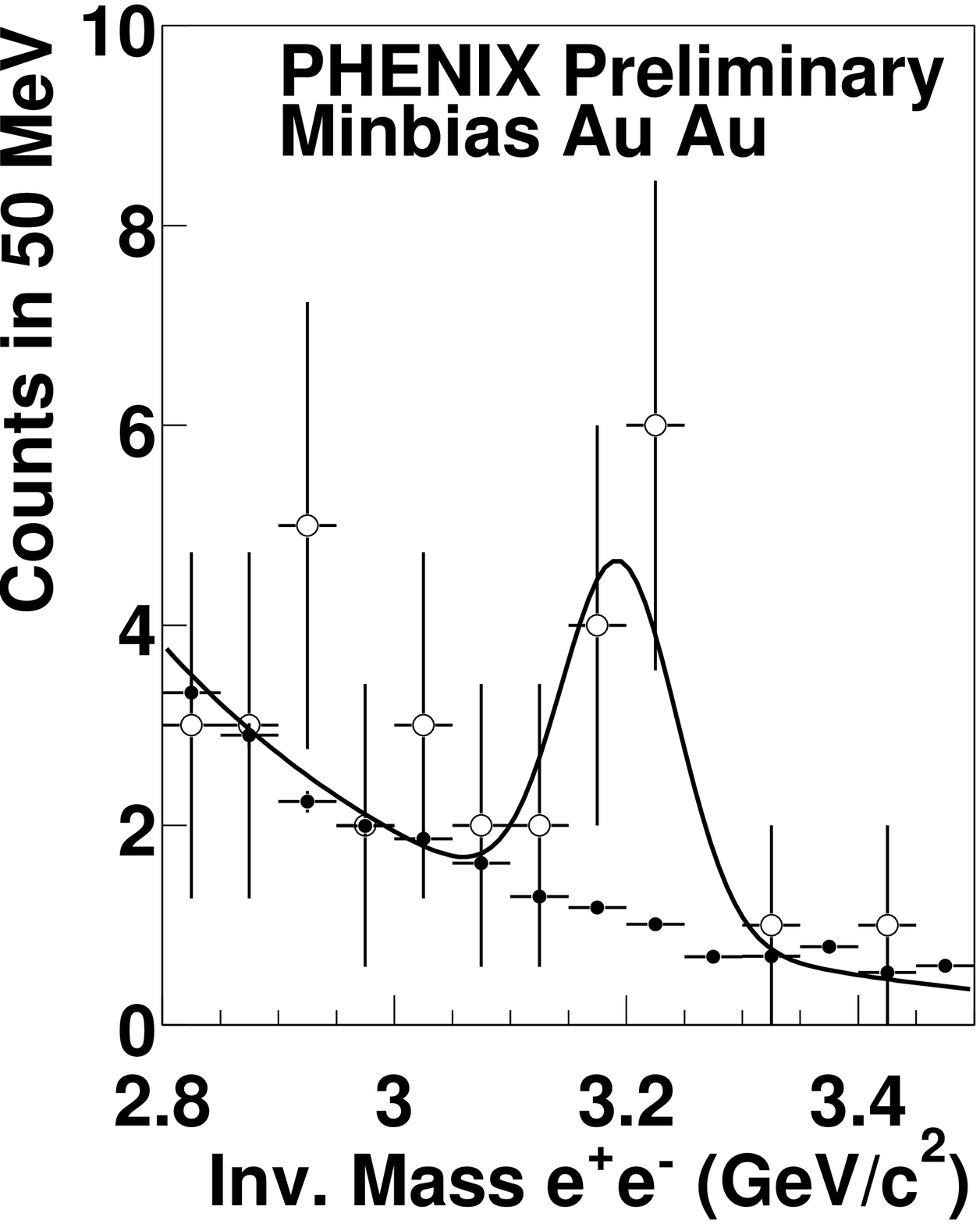}
\end{minipage}\\
\begin{minipage}[t]{0.3\linewidth}
\centering
\caption{Dielectron invariant mass distribution from proton-proton events 
where at least one 2x2 EMC tile is above our Level-1 trigger threshold.}
\label{fig:jpsi_pp_ee}
\end{minipage}%
\hspace{0.2in}
\begin{minipage}[t]{0.3\linewidth}
\centering
\caption{Dimuon invariant mass distribution (points), with
unlike sign pair background (histogram) from proton-proton events
that fired the Muon Identifier Level-1 dimuon trigger.}
\label{fig:jpsi_pp_mumu}
\end{minipage}%
\hspace{0.2in}
\begin{minipage}[t]{0.3\linewidth}
\centering
\caption{Invariant mass $e^{+}e^{-}$ distribution for "minimum bias" (0-90\% central) 
Au-Au collisions at $\sqrt{s_{NN}}=200$ GeV.  Also shown is the mixed event background 
distribution.}
\label{fig:jpsi_auau_mb_mass}
\end{minipage}
\end{figure}

At this point, we can make the following qualitative observations.  
(1) Our electron data are consistent with charm expectations and binary scaling within our current 
statistical and  systematic errors. 
(2) The NA50 experiment at the CERN-SPS has observed an enhancement in 
intermediate mass dimuons~\cite{na50-charm}.  One possible interpretation is a factor of three
enhancement of charm in central Pb-Pb collisions relative to proton-proton. 
We do not see such an effect at RHIC energies.
(3) PHENIX reports a substantial suppression in high transverse momentum
$\pi^{0}$ relative to binary scaling~\cite{highpt}.  We do not see such a large suppression
in the single electrons from charm.  This could be due to a lower charm quark 
energy loss in the dense gluonic medium due to the ``dead-cone'' effect.
Further quantification of these observations is forthcoming.

\section{J/$\psi$ Physics}

We expect a screening of the QCD attractive potential as we approach the
deconfinement transition.  The color screening may result in a decrease in
the production of heavy quarkonia ($J/\psi, \psi', \Upsilon$, etc.)~\cite{satz}.
Alternatively, there are models that predict an enhancement of heavy quarkonia
due to $c\overline{c}$ coalescence as the collision volume cools~\cite{thews}.  
This enhancement has a large dependence on the total charm production.

\subsection{Proton-Proton Results}

We present here the first measurements of $J/\psi$ production in proton-proton
collisions at $\sqrt{s}=200$~GeV.  We show in Figure~\ref{fig:jpsi_pp_ee} the
dielectron invariant mass distribution for proton-proton collisions with
at least one 2x2 calorimeter tower tile above our Level-1 trigger threshold.
This represents about half of our proton-proton
statistics, some of which are from additional 4x4 calorimeter tower tile
Level-1 triggers.
We show in Figure~\ref{fig:jpsi_pp_mumu} the dimuon invariant mass distribution
for proton-proton collisions.

\begin{figure}[htb]
\begin{center}
\includegraphics[width=20pc]{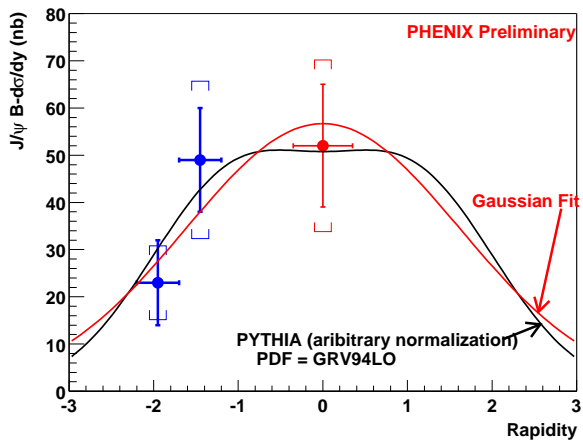}
\caption{$B_{ll}~d\sigma/dy$ as a function of rapidity for $pp \longrightarrow J/\psi + X$.}
\label{fig:jpsi_rap}
\end{center}
\end{figure}

We have corrected for the central and muon arm acceptances, tracking
and particle identification efficiencies, and Level-1 trigger efficiencies.
We show in Figure~\ref{fig:jpsi_rap} the $J/\psi$ branching fraction to leptons (B)
times $d\sigma/dy$ as a function of rapidity.  The central
rapidity point is from our dielectron measurement and the muon data is
divided into two rapidity bins.  The data are fit to a simple Gaussian form and
also to the predicted rapidity shape from PYTHIA using parton distribution
function GRV94LO.  Removing the branching fraction and integrating over
rapidity, we find:
\begin{equation}
\sigma(pp \longrightarrow J/\psi+X) = 3.8 \pm 0.6({\rm stat}) \pm 1.3({\rm sys})~~\mu{\rm b}~~~{\rm [PRELIMINARY]}
\end{equation}

\subsection{Gold-Gold Results}

\begin{figure}[htb]
\begin{center}
\includegraphics[height=18pc]{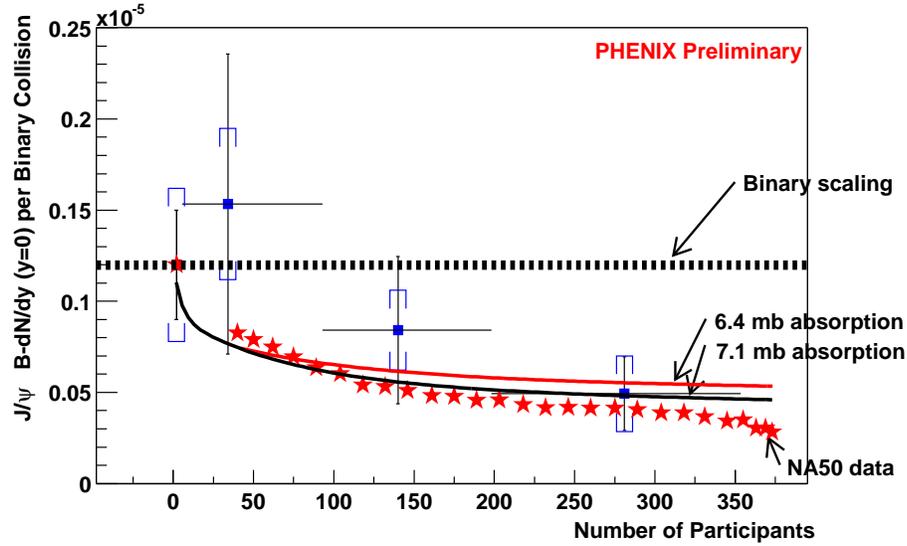}
\caption{$J/\psi$ branching fraction to electrons times dN/dy at midrapidity (y=0) as
a function of collision centrality (Number of Participants).  The yields are per
calculated binary collision.  Shown are both the PHENIX proton-proton and Au-Au data points.}
\label{fig:jpsi_scaling_models}
\end{center}
\end{figure}

Shown in Figure~\ref{fig:jpsi_auau_mb_mass} is the dielectron invariant
mass distribution for ``minimum bias'' (0-90\% central) Au-Au collisions, along
with a mixed event distribution.  Further details are in~\cite{tony}.
Although the statistics are limited, we have calculated $B_{ee}~dN/dy|_{y=0}$
in three Au-Au centrality ranges, where $B_{ee}$ is the $J/\psi$ branching fraction to
electrons.
The ranges correspond to 0-20\%, 20-40\%, and 40-90\% central with calculated
average number of binary collisions of 791, 297, and 45, respectively.  We show the 
$B_{ee}~dN/dy|_{y=0}$ per binary collision as a function of the collision centrality (represented by
the number of calculated participating nucleons) in Figure~\ref{fig:jpsi_scaling_models}.
Note that we have also included our midrapidity $J/\psi$ measurement in proton-proton collisions with 
one binary collision.  The statistical errors are shown as lines and the systematic errors
as brackets.  

On the same figure, we show different possible models of the $J/\psi$ centrality dependence.  
All the models
have been arbitrarily normalized to perfectly intersect the center of our proton-proton data point.
This graphical presentation can be somewhat misleading since it effectively assumes no error on the
proton-proton measurement.  However, we can perform an unbiased measure of agreement 
using a statistical confidence level.  
The statistical confidence level is defined as the probability,
based on a set of measurements, that the actual probability of an event is better than some
specified level.  We find that the statistical confidence level is 16\%, 80\%, and 75\% for (1)
a model assuming binary collision scaling of the $J/\psi$, (2) a model assuming binary collision
scaling followed by ``normal'' nuclear absorption with a 7.1 mb cross section, and (3)
the exact pattern of suppression relative to binary scaling observed by the NA50 experiment
at lower energies~\cite{na50-jpsi}, respectively.  It is notable that although the binary collision
scaling model is statistically less favorable, all three models have acceptable confidence levels
without consideration of our current systematic errors.
Also, we hope to measure the ``normal'' nuclear absorption in deuteron-nucleus reactions at RHIC 
in the next two years.

There are two extreme models of $J/\psi$ production at RHIC.  One is a 
simple extrapolation from models used to describe the lower energy NA50 results that imply that 
almost all $c \overline{c}$ correlations are broken up and one should have over an order of magnitude
more suppression~\cite{zhang}.  Another is that although all initial $J/\psi$ are destroyed,
they are re-created via coalescence of uncorrelated $c\overline{c}$ in the late stages of the time
evolution.  This mechanism can lead to an enhancement of $J/\psi$ relative to binary collisions (which
is statistically less likely given our data).  

With the data set from Run II, we can expect a factor of two more statistics from our Level-2 
dielectron triggered data in Au-Au, in addition to results at forward rapidity 
from our dimuon measurements.
Also, we are now confident in our abilities to trigger at high RHIC luminosity, and eagerly 
await a larger data set.

\section{Summary}

In summary, the PHENIX program of leptonic observables has started with data from Run II.
We have the first observation of $J/\psi$ at $\sqrt{s_{NN}}=200$ GeV in both proton-proton and
Au-Au systems.  We have shown results on $\phi$ production, dielectron continuum, and
single electrons from charm D meson decay.  
We have shown just the tip of the iceberg of the PHENIX leptonic observables, and 
are excited about prospects for future
high luminosity running at RHIC.

%%%%%%%%%%%%%%%%%%%%%%%%%%%%%%%%%%%%%%%%%%%%%%%%%%

\end{document}